## Journal of Geophysical Research: Atmospheres

# Constraints to do realistic modeling of the electric field ahead of the tip of a lightning leader

**Alexander Broberg Skeltved[1]**, **Nikolai Østgaard[1]**, **Andrew Mezentsev[1]**, **Nikolai Lehtinen[1]**, and **Brant Carlson[1,2]**

[1]Birkeland Centre for Space Science, Institute of Physics and Technology, University of Bergen, Bergen, Norway, [2]Physics and Astronomy, Carthage College, Kenosha, Wisconsin, USA

**Abstract** Several computer models exist to explain the observation of terrestrial gamma-ray flashes (TGFs). Some of these models estimate the electric field ahead of lightning leaders and its effects on electron acceleration and multiplication. In this paper, we derive a new set of constraints to do more realistic modeling. We determine initial conditions based on in situ measurements of electric field and vertical separation between the main charge layers of thunderclouds. A maximum electric field strength of 50 kV/cm at sea level is introduced as the upper constraint for the leader electric field. The threshold for electron avalanches to develop of 2.86 kV/cm at sea level is introduced as the lower value. With these constraints, we determine a region where acceleration and multiplication of electrons occur. The maximum potential difference in this region is found to be ~52 MV, and the corresponding number of avalanche multiplication lengths is ~3.5. We then quantify the effect of the ambient electric field compared to the leader field at the upper altitude of the negative tip. Finally, we argue that only leaders with the highest potential difference between its tips (~600 MV) can be candidates for the production of TGFs. However, with the assumptions we have used, these cannot explain the observed maximum energies of at least 40 MeV. Open questions with regard to the temporal development of the streamer zone and its effect on the shape of the electric field remain.

## 1. Introduction

Terrestrial gamma-ray flashes (TGFs) are short, intense, very energetic bursts of bremsstrahlung photons that are produced by relativistic electrons. Two leading scenarios have been presented to explain how free electrons can be accelerated to relativistic energies and multiplied in the thundercloud electric fields.

1. *Wilson* [1925] presented the idea that high-energy electrons, such as cosmic ray secondaries, in the presence of a strong electric field can be accelerated to overcome the friction force in air. These electrons then become runaway electrons. It is well established that the ambient field in thunderclouds can be sufficiently strong for this to occur. *Gurevich et al.* [1992] later proposed that runaway electrons can undergo further multiplication, primarily through the Møller scattering process, producing Relativistic Runaway Electron Avalanches (RREAs). In addition to high-energy electrons, also, positrons and high-energy photons will be created and some will backscatter and create seed particles for new RREAs [*Dwyer*, 2003, 2012; *Skeltved et al.*, 2014]. This multiplication of RREAs is called the feedback mechanism. Thus, this scenario is explained by the ability of the ambient field to accelerate electrons enough to overcome the friction force of air over the vertical distance between the main charge layers in thunderclouds.

2. *Moss et al.* [2006] proposed that acceleration and multiplication of electrons can occur in the strong inhomogeneous electric field created ahead of a lightning leader. They also suggested that seed electrons in this scenario are thermal electrons accelerated in streamer tips during the very early stage of streamer development. Modeling results suggest that the electric fields in streamer tips are indeed sufficient to accelerate electrons to average energies of ~65 keV [*Celestin and Pasko*, 2011]. Several studies have reported modeling results of this scenario and concluded that it may be sufficient to explain the number of electrons and gamma rays that has been observed [*Carlson et al.*, 2010; *Celestin and Pasko*, 2011; *Xu et al.*, 2012; *Köhn and Ebert*, 2015]. The effect of feedback in the leader field has also been discussed [*Dwyer*, 2007; *Carlson*, 2009] and was shown to depend on the geometry and the charge density in the tip of the leader. Results from *Köhn et al.* [2017] indicate that the effect of feedback could be important depending on the strength and shape of the leader electric field.

**Table 1.** A Comparison of the Two Scenarios and the Role of the Important Mechanisms[a]

| Scenarios | Runaway Electrons | RREAs | Feedback |
|---|---|---|---|
| (1) Ambient field | Any, including thermal | Yes | Yes |
| (2) Leader field | Thermal | Yes | Yes (< ambient) |

[a]Note that the primary difference is the electric field configuration, which is uniform in scenario 1 and highly nonuniform in scenario 2. In addition, although feedback has been shown to be present in both models, the effect is likely larger for scenario 1 than for scenario 2.

Notice that there are two primary differences between the scenarios. The runaway electrons of scenario 1 can be supplied from any source of above ~10 keV, whereas scenario 2 relies on the acceleration of thermal electrons. The electric field is relatively uniform and extended to kilometer scale in scenario 1 but highly nonuniform and localized to hundred meter scale in scenario 2. In both scenarios, extreme thundercloud conditions have been assumed to fully account for the observations [*Dwyer*, 2003; *Xu et al.*, 2012; *Köhn and Ebert*, 2015; *Celestin et al.*, 2012]. Table 1 summarizes the similarities and differences between the two scenarios. In reality, when both the ambient field and local enhancements exist, the question is as follows: what is their relative importance for TGF production?

The purpose of this paper is to establish realistic constraints to derive the effects of the leader electric field on the acceleration and multiplication of high-energy electrons. We first discuss observable properties of TGFs then present a description of relevant properties of positive intracloud leader (+IC). We go on to present the initial conditions of existing models and how these models estimate the leader electric field. It will then be shown that the strength and shape of the estimated electric fields depend largely on the method used to calculate them. This result will be shown to have important implications for the region where acceleration and multiplication of high-energy electrons can occur. We then present the assumption that the positive end of the channel branches and develops horizontally in the negative charge region. The system of branched channels can than be approximated as a partly conducting plane relative to the negative end of the channel. Effectively, this explains how the potential difference between a point just ahead of the leader and the ambient potential can be increased, but not more than doubled. Finally, we use the new constraints to make estimates of the maximum energy and multiplication rate of high-energy electrons.

## 2. Properties of TGFs and of Their Source Electrons

An extensive overview of TGF properties and related phenomena is found in *Dwyer et al.* [2012]. In this section we summarize the parameters that are important from the perspective of the production mechanisms.

TGFs are submillisecond bursts of photons with maximum energies of up to at least 40 MeV [*Smith et al.*, 2005; *Briggs et al.*, 2010; *Marisaldi et al.*, 2010]. Comparisons between modeling results and the average photon energy spectrum obtained from satellite measurements have indicated that TGFs originate from inside thundercloud regions in the Earth's atmosphere at altitudes below 21 km [*Dwyer and Smith*, 2005; *Carlson et al.*, 2007; *Østgaard et al.*, 2008; *Gjesteland et al.*, 2010]. Measurements of radio waves produced by lightning and their association with TGFs have indicated that the typical production altitude is actually between 10 km and 15 km altitude [*Cummer et al.*, 2005, 2011, 2015; *Gjesteland et al.*, 2015; *Shao et al.*, 2010; *Lu et al.*, 2010]. The energy distribution of the photons has been explained by an attenuated bremsstrahlung spectrum indicating that the source are high-energy electrons [*Lehtinen et al.*, 1996]. *Skeltved et al.* [2014] also found that the ratio of high-energy (>100 keV) electrons to photons is between 1 and 10. Finally, better search algorithms [*Østgaard et al.*, 2012, 2015] and new detector configurations [*Marisaldi et al.*, 2015] have shown that TGFs are more common than previously thought.

To reach the numbers and energies that have been inferred from measurements, high-energy (>10 keV) electrons must be present in a region with a strong electric field. Such electrons are thought to be thermal electrons that have been accelerated in the strong electric field near the tips of streamers [*Moss et al.*, 2006] or the product of cosmic ray secondary particles [*Wilson*, 1925]. Note that thermal electrons refer to the seed electrons that have been accelerated from energies in the eV range to become runaway electrons in the tens of keV range; this process can also be referred to as thermal acceleration. Further multiplication is dominated by elastic Møller scattering of electrons in electric fields stronger than the RREA threshold, $E_{RREA} = 2.86$ kV/cm at sea level [Coleman and Dwyer, 2006]. The RREA threshold is slightly above the minimum stopping power of air, since the electrons do not go in exactly a straight line. This multiplication process is the RREA mechanism and can be described by the following equation [Gurevich et al., 1992; Gurevich and Zybin, 2001]:

$$N_{RREA} = N_o \cdot e^{L/\lambda(E)} \tag{1}$$

where $N_{RREA}$ is the total number of electrons from the multiplication of $N_o$ seed electrons, $L$ is the length of the multiplication region, and $\lambda(E)$ is the avalanche (e-folding) length in a homogeneous electric field, $E$. The derivation of equation (1) can be found in Gurevich et al. [1992] and Skeltved et al. [2014]. For electric field strengths between the conventional breakdown threshold, $E_{th} = 32$ kV/cm, and the RREA threshold $E_{RREA}$, the avalanche length is given by the following two functions

$$\lambda(E) \approx \frac{7.3 \times 10^3 \text{kV}}{E - 2.75 \text{kV/cm}} \tag{2}$$

in the range $32$ kV/cm $> E > 3$ kV/cm and

$$\lambda(E) \approx \frac{5.1 \times 10^3 \text{kV}}{E - 2.85 \text{kV/cm}} \tag{3}$$

in the range $3$ kV/cm $> E > 2.86$ kV/cm. These functions are based on modeling results [Dwyer, 2003; Coleman and Dwyer, 2006] and have been validated by Dwyer [2012] and Skeltved et al. [2014]. The total number of avalanche lengths $N_\lambda$ in a uniform electric field $E$, over a distance $L$, is then

$$N_\lambda = \frac{L}{\lambda(E)}. \tag{4}$$

In a strongly nonuniform electric field, such as that ahead of a lightning leader (see section 3), the electric field, $E(l)$, depends on the distance, $l$. The total number of avalanche lengths then becomes

$$N_\lambda = \int_{l_1}^{l_2} \frac{dl}{\lambda(E(l))}, \tag{5}$$

where $\lambda(E(l))$ is defined by equations (2) and (3).

In order to produce the observed bremsstrahlung spectrum of TGFs, the energy distribution of the RREAs must reach a steady state with maximum energies greater than that of TGFs, which is up to at least 40 MeV [Marisaldi et al., 2010].

## 3. The +IC Leader and the Effects of Its Electric Field

When a lightning leader develops in the ambient electric field of a thundercloud, it creates its own strong electric field ahead of its tip. To understand how this may be important for the production of TGFs, a short description of positive intracloud (+IC) leaders will be given.

The thundercloud can be approximated by three vertically stacked charge layers [Williams, 1989]. A weak lower positive charge layer, a strong main negative (MN) charge layer, and a slightly weaker upper positive (UP) charge layer. The MN charge layer is thin, with a higher charge density [Marshall and Stolzenburg, 2001; Rakov and Uman, 2003] and is centered where the temperature is between −10° and −25°C, typically from 4 km to 6 km altitude [Rakov and Uman, 2003, p. 75]. The UP charge layer is less dense, but more vertically extended, and is typically centered at altitudes between 10 km and 14 km [Rakov and Uman, 2003, p. 76]. The maximum altitude of thunderclouds is determined by the height of the tropopause that, depending on altitude and season, can reach 17 km [e.g., Pan and Munchak, 2011].

The amount of charge in the charge layers and the separation between them determines the strength of the ambient electric field. Marshall and Rust [1991], Marshall and Stolzenburg [1996, 2001], and Stolzenburg et al. [2007] have presented electric field soundings through a range of thunderstorms. From their measurements it can be seen that the mean strength of the ambient electric field between the main charge regions is roughly between 0.2 kV/cm and 0.5 kV/cm. These values are consistent with other studies [*Rakov and Uman*, 2003; *Coleman*, 2003]. In *Marshall and Stolzenburg* [2001], cloud tops where measured to be from 9 to 12.5 km and the gap between the main charge regions extended between 2 and 5 km vertically. The negative charge layer was located between 5 km and 7 km altitude. These values are close to the typical values from *Rakov and Uman* [2003]. Although the peak strength of the ambient electric field can reach values stronger than the RREA threshold, the mean values are typically weaker [*Stolzenburg et al.*, 2007]. Thus, for altitudes closer to 14 km the mean ambient electric field can be expected to be less than the RREA threshold $E_{RREA}(14) = 0.48$ kV/cm. By integrating the electric field over the vertical extension, the potential difference between the charge layers can be found. *Marshall and Stolzenburg* [2001] investigated the potential differences, also using electric field soundings through thunderstorms. The largest potential difference they found was around 130 MV and occurred between 6.5 and 9.9 km altitude.

IC leaders are initiated when the electric field exceeds the conventional breakdown threshold, $E_{th} = 32$ kV/cm at sea level. Typically, this can occur in a localized region just above the MN charge region. The IC leader is characterized as a highly conductive plasma channel that develops to a vertical length, $L$, of a few kilometers and has a radius, $r$, of approximately 1 cm [*Cooray*, 2015, p. 250], sea level equivalent. IC leaders are bidirectional, with charge of opposite polarity concentrated at the two ends of the channel. As the channel develops in the ambient electric field of the cloud, the amount of charge in each end increases. However, the total charge on the leader remains zero [*Kasemir*, 1960]. After initiation, the leader develops until it spans the gap between the two charge layers. The channel is then discharged, sometimes in several consecutive discharge processes, where also branching of the channels in the charge regions can occur. The IC lightning, as described here, is believed to be the most common type of lightning flash [*Rakov and Uman*, 2003, p. 321].

TGFs have been associated with radio signals generated by lightning discharges [*Inan et al.*, 1996]. In particular, a close relationship between TGFs and +IC lightning, which develop in a stepped manner transporting electrons from the MN charge layer toward the UP charge layer, has been reported [*Cummer et al.*, 2005; *Stanley et al.*, 2006; *Shao et al.*, 2010; *Lu et al.*, 2010; *Cummer et al.*, 2011; *Østgaard et al.*, 2013; *Cummer et al.*, 2015].

If one approximates the leader channel as perfectly conducting, the induced charge will distribute itself at the surface of the channel to exactly oppose the ambient electric field $E_o$. The electric potential is then uniform through the center of the leader and accumulates ahead of the tips. In reality there will be a small potential drop along the channel due to the finite conductivity. This is not taken into account in this study. The potential difference, $\Delta U_{tip}$, between the tip of a perfectly conducting channel and the ambient potential, is equal to the potential difference from the center of the channel to the tips [*Bazelyan and Raizer*, 2000, p. 54], which is given as

$$\Delta U_{tip} = \frac{1}{2} \int_{h_o}^{h_1} E_o(h) dh = \frac{E_o^{av} \cdot L}{2}, \tag{6}$$

where $h_o$ and $h_1$ are the lower an upper altitudes of the channel and $E_o(h)$ is the ambient electric field at altitude $h$. This approximation is consistent with previous studies [*Celestin and Pasko*, 2011; *Xu et al.*, 2012, 2015; *Celestin et al.*, 2015; *Mallios et al.*, 2013; *Pasko*, 2013; *Köhn and Ebert*, 2015].

In the region ahead of the tips, where the electric field is above the conventional breakdown threshold $E_{th}$, streamers are continuously initiated. Streamers are small plasma channels of low conductivity that carry a high density of electrons at their tip and leave behind ions and free electrons in their wake. At the positive end of the leader, positive streamers are created by electron avalanches directed toward the streamer tips (due to the polarity of the electric field). The positive channel therefore develops continuously at a speed of roughly $\sim 10^6$ m/s. At the negatively charged leader tip, the negative streamers are created by electron avalanches propagating outward, leading to a stepped development that is an order of magnitude slower (on average $\sim 10^5$ m/s). During the outward expansion of the streamers, a leader stem typically forms at a distance of between several tens [*Rakov and Uman*, 2003] to hundreds [*Cummer et al.*, 2015] of meters ahead of the channel. The formation of the leader stem is still poorly understood. The leader stem is also bidirectional, with a positive channel that develops toward the main leader tip and a negative end that develops in the opposite direction. Both are preceded by smaller streamer zones of their own. When the positive secondary channel attaches to the main channel, a discharge wave redistributes the charge on the leader channel such that the tip of the former leader stem becomes the new leader tip [*Bazelyan and Raizer*, 2000, pp. 197–198]. How fast the potential is increased on the new tip is not known.

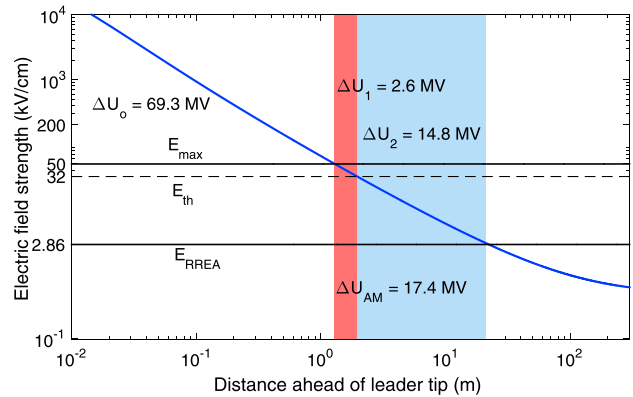

**Figure 1.** The electric field of a capped channel of 4 km length, 1 cm radius immersed in an ambient field of 0.5 kV/cm. The solid black lines indicate the maximum field at the leader tip, $E_{max} = 50$ kV/cm, and the RREA threshold, $E_{RREA} = 2.86$ kV/cm. The dashed line indicates the conventional breakdown threshold, $E_{th} = 32$ kV/cm, sea level equivalent. The colored areas cover the AM region, red for $E_{max} > E \geq E_{th}$ and blue for $E_{th} \geq E \geq E_{RREA}$ (equations (2) and (3)). $\Delta U_0$, $\Delta U_1$, and $\Delta U_2$ denote the potential differences in the very high field region $E > E_{max}$ and the AM regions, respectively.

Modeling results have indicated that the electric field at the tip of streamers can reach values of ~10 $E_{th}$, which thereby is one of the unique circumstances when low-energy electrons can be accelerated to runaway energies [*Moss et al.*, 2006]. *Celestin and Pasko* [2011] estimated that roughly one out of a hundred low-energy electrons can be accelerated to an average energy of 65 keV. *Moss et al.* [2006] also proposed that further acceleration can occur in the weaker but more extended leader electric field. The presence of streamers and thus of charged particles increases the conductivity in the streamer zone significantly and thereby limits the strength of leader electric field. As is explained in *Bazelyan and Raizer* [2000, p. 68], the electric field cannot exceed the ionization threshold by much, because any excess charge will quickly reduce the field and stabilize it closer to the conventional breakdown threshold $E_{th}$. They also discussed laboratory experiments that have shown the maximum field, $E_{max}$, at the tip of a leader channel with radius ~1 cm to be roughly $1.5 \cdot E_{th} \sim 50$ kV/cm. The maximum electric field that can exist ahead of a lightning leader is therefore likely to depend on how quickly the potential on the new tip increases [*Celestin and Pasko*, 2011] and on how fast screening due to ionization occurs. It is generally believed that after the initial stage of potential transfer, called the corona flash stage, the electric field in the streamer zone stabilizes close to the field required to sustain negative streamer development, which is ~12.5 kV/cm at sea level [*Moss et al.*, 2006].

### 3.1. Introducing Electric Field Constraints

Similar to previous models, we consider a static picture of the leader, when all the potential has been transferred to the new tip [*Celestin and Pasko*, 2011; *Celestin et al.*, 2012, 2015; *Xu et al.*, 2012, 2015; *Köhn and Ebert*, 2015]. To create a self-consistent model that includes the effects of a developing streamer zone on the electric field is beyond the scope of this study. However, to avoid unrealistically strong electric fields, we limit the maximum electric field strength to an upper limit of $E_{max} = 50$ kV/cm, at sea level density and pressure. Note that the screening of the electric field inside the streamer zone may also increase the electric field at the edge and thereby further complicate the system. This effect has not been considered. This limit is in accordance with the arguments presented in the previous section and with *Bazelyan and Raizer* [2000, p. 68], *Celestin and Pasko* [2011], and *Celestin et al.* [2015]. Furthermore, we assume that the seed electrons are accelerated from eV energies to keV energies in the tips of streamers [*Celestin and Pasko*, 2011]. The initial position of these seed electrons, after acceleration, is then assumed to be at the position that corresponds to the upper boundary of the electric field, $E_{max}$. In order for RREAs to be sustained, the electric field strength must also exceed the RREA threshold, $E_{RREA} = 2.86$ kV/cm at sea level [*Gurevich et al.*, 1992; *Dwyer*, 2003], below which even the high-energy particles (>1 MeV) will quickly stop. Thus, the region ahead of the leader tip where the electric field falls from $E_{max}$ to $E_{RREA}$ will be called the Acceleration and Multiplication (AM) region hereafter. In Figure 1, we present the sum of the electric field created by the leader and the ambient electric field (see section 4 for description of the leader electric field). The horizontal solid and dashed lines indicate the electric field thresholds, and the AM region is shown as the red and blue areas, where the red area indicates electric fields between $E_{max} > E > E_{th}$ and the blue area indicate the electric fields between $E_{th} > E > E_{RREA}$ (equations (2) and (3)). If the electric field created by the leader is $E(l)$ (see description in section 4) and the ambient electric field is $E_o$ at the upper altitude of the leader, the potential difference $\Delta U$ in a region ahead of the leader can then be calculated by

$$\Delta U = \int_{l_1}^{l_2} E(l)dl + E_o \cdot (l_2 - l_1).$$ (7)

where $l_1$ and $l_2$ are the distances to the boundaries of the region.

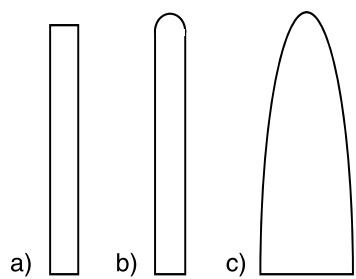

**Figure 2.** The geometries used to approximate the leader channel. (a) A long flat-ended channel. (b) A channel with caps. (c) An ellipsoid. Note that the sketch is not to scale and only the upper half of the channel is depicted.

## 4. Existing Computer Models

In the preceding sections we have described how the strong electric field created by +IC leaders may produce TGFs. We will now describe the estimated electric fields and their corresponding potential differences and present the initial conditions that have been used in earlier studies. In this context, there are four important issues to consider: (1) the geometry of the leader channel (shape of the tip, vertical length, and radius); (2) the ambient electric field, which combined with the length also defines the potential difference between the tips of the leader channel; (3) the upper altitude of the leader, $h$; and (4) how close to the leader tip the seed electrons can be initiated.

To approximate the leader channel, it is common to consider a perfectly conducting object immersed in an ambient electric field. In existing studies this is usually either a long thin wire with flat ends [*Celestin and Pasko*, 2011; *Xu et al.*, 2012; *Celestin et al.*, 2012; *Xu et al.*, 2015; *Mallios et al.*, 2013; *Pasko*, 2013; *Celestin et al.*, 2015] or an ellipsoid with a curvature radius of 1 cm at the tip [*Köhn and Ebert*, 2015]. In order to avoid confusion between a physical wire and the leader channel that may act similar to a wire, we will use the term channel throughout this paper. The ellipsoid has the advantage of having an analytical solution [*Landau and Lifshitz*, 1960, pp. 20–27] and also the disadvantage of a geometry less similar to a natural leader channel. For the flat-ended channel the numerical method of moments has been used [*Balanis*, 2012, pp. 679–691]. See Appendix A for the general method of solving for the charge distribution on a cylindrical symmetric conducting object. Using this charge distribution, one can calculate the electric field, $E$, created by the leader at an arbitrary point in space:

$$E(\mathbf{r}) = \frac{1}{4\pi\epsilon_0} \int_S \frac{\sigma(\mathbf{r}')ds}{|\mathbf{r},\mathbf{r}'|^2},$$ (8)

where $\sigma$ is the calculated charge density, integration is over the conductor surface, and $\mathbf{r}$ is a point on the axis ahead of the channel tip. In this paper, we use the stable "surface" method of moments. In this approach, one chooses both the observation points and source points on the surface [*Harrington*, 1993, pp. 28–33], when estimating the charge distribution, as opposed to having observation points on the center axis of the channel [*Balanis*, 2012, pp. 679–691]. In Appendix A, section A2, we describe both methods. Contrary to the latter [*Balanis*, 2012], the first approach [*Harrington*, 1993] gives a stable solution for element sizes smaller than the radius of the channel. The stability of the two methods is discussed in Appendix A, section A3. The results presented in this paper will be calculated for a channel with capped tips. The caps will be half spheres with radii equal to the radius of the channel in order to make the geometry at the channel tip more realistic. Note that we also use the stable method of moments to solve for the flat-ended channel in section 5.1. The geometries at the tips of the channels are illustrated in Figure 2.

*Celestin and Pasko* [2011] used an ambient electric field of 0.2 kV/cm and a length of 1 km, giving a potential difference at the tip, between the leader and the ambient potential, of $\Delta U_{tip} = 10$ MV (equation (6)). The model by *Celestin and Pasko* [2011] is capable of simulating the acceleration of electrons to runaway energies but does not include a full-scope simulation of the production of TGFs. *Xu et al.* [2012, 2015] used a 4 km lightning leader immersed in an ambient field of 0.5 kV/cm, which corresponds to a total potential difference over the vertical length of the leader of 200 MV and a potential difference at the tip $\Delta U_{tip} = 100$ MV. They initiate the electrons at 30 cm from the tip, where we calculate the electric field to be 263.3 kV/cm based on the leader parameters from *Xu et al.* [2012] but using the stable method of moments. The maximum field corresponds to roughly 8 times the conventional breakdown threshold and 5 times the maximum field strength described in *Bazelyan and Raizer* [2000, p. 68], at sea level. *Köhn and Ebert* [2015] used the same parameters as *Xu et al.* [2012] but calculated the electric field using the analytical solution of an ellipsoid. They also initiated the electrons at 30 cm from the leader tip, which corresponds to a stronger electric field of $E_{max} = 283.0$ kV/cm due to the geometry. In addition, they perform their simulations at 16 km altitude, which lead to a significantly higher maximum field to breakdown threshold ratio $E_{max}/E_{th}(16) \approx 74$, which is rather extreme. In this electric field, they are able to simulate the acceleration of electrons from 0.1 eV to tens of MeV. Note also that an ambient electric field of 0.5 kV/cm at 16 km altitude corresponds to ~1.45 · $E_{RREA}$, which is also much larger than typical measurements suggest. *Celestin et al.* [2015] considered five cases with potential differences between the leader tips that range from 10 MV to 600 MV, which is 5 MV to 300 MV from the center to the tip. The length of the channels was assumed to be from 1 km to 6 km, and they used the same constraints on the electric field that we present in this paper. In the case presented by *Celestin et al.* [2012], they used a leader length of 3.5 km and an ambient electric field of 2.0 kV/cm, which gives a total potential difference over the vertical length of the leader of 700 MV and $\Delta U_{tip} = 350$ MV ahead of its tip. They initiated the seed electrons as close as 15 cm from the leader tip, where we calculate the electric field to be 2152.4 kV/cm (using the stable method of moments), which is roughly 60 times conventional breakdown threshold at sea level. Note that in *Celestin et al.* [2012], they acknowledged that the assumed electric fields are rather extreme.

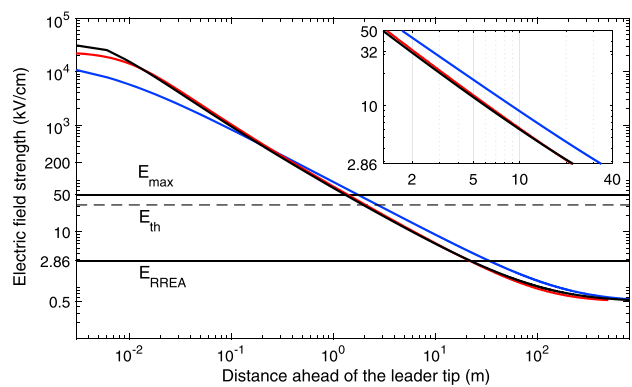

**Figure 3.** A comparison of the electric field calculated along the center axis ahead of a perfectly conducting channel of 4 km length and 1 cm radius immersed in an ambient electric field of 0.5 kV/cm. The capped cylinder is indicated by the black line, the flat-ended channel is red, and the ellipsoid is blue. The stable method of moments has been used for the red and black curve. The solid horizontal lines correspond to the threshold field strengths $E_{max}$ and $E_{RREA}$, and the dashed line is the conventional breakdown threshold $E_{th}$. In the smaller panel, we have zoomed in on the AM region, the range between $E_{max}$ and $E_{RREA}$. Note that although the field is directed toward the leader tip, the field strength is presented as positive to make the figure more presentable.

## 5. Results

### 5.1. Dependence of the Electric Field on the Channel Shape

Different methods have been used to estimate the electric field ahead of the leader channel. In Figure 3 we show the electric field calculated for the conducting ellipsoid, the flat-ended channel, and the capped channel with a length of 4 km. The electric field is calculated by using the stable method of moments for the flat-ended and capped channels, and the analytical method is used for the ellipsoid. The flat-ended channel and capped channel have a radius of 1 cm both along the channel and at the spherical caps. The ellipsoid has a fixed curvature radius of 1 cm at the tip. The ambient field strength is chosen to be 0.5 kV/cm, which corresponds to a total potential difference over the entire channel of 200 MV and 100 MV from its center (equation (6)). In addition, the two horizontal lines indicate $E_{max} = 50$ kV/cm and $E_{RREA} = 2.86$ kV/cm (sea level), and the dashed line indicates the conventional breakdown threshold, $E_{th} = 32$ kV/cm, sea level equivalent. In the smaller panel, we zoom in on the AM region between $E_{max}$ and $E_{RREA}$.

From this comparison, we can see that the geometry has an important effect on how the electric field changes with distance from the tip of the leader. The ellipsoid has a much larger radius at the center of the channel. This results in a weaker maximum electric field compared to the flat-ended channel and capped channel geometry and also a more extended region where the field is above the RREA threshold. The tip of the capped channel is the most pointed and produces the strongest electric field, which is within roughly 1 cm of the tip. Most importantly, we have calculated the electric potential in the AM region to be roughly 26.6 MV for the ellipsoidal channel and 17.4 MV for the flat-ended and capped channels, respectively. However, the ellipsoid has an unrealistic radius at the center of the channel of 6.3 m and we will argue that the resulting shape of the electric field may also be unrealistic. The difference between the flat-ended channel and capped channel is negligible in the range of the AM region.

### 5.2. Potential Differences and Number of Avalanche Lengths in the AM Region

The initial conditions that are used in this paper are within the range observed for typical [+]IC lightning and TGFs. These values are presented in Table 2, which gives the potential difference between the leader and the ambient potential at the tips, $\Delta U_{tip}$. These are given by assumptions of different lengths, $L$, altitudes for the upper leader tip, $h$, and ambient field strengths, $E_o$ (calculated using equation (6)). The initial conditions are within the range of in situ measurements presented by *Marshall and Stolzenburg* [2001] and *Stolzenburg et al.* [2007]. Note that in the fourth row, we give the range of altitudes for +IC leaders associated with the production of TGFs in *Shao et al.* [2010] and *Lu et al.* [2010]. Here we use the maximum vertical separation and an ambient electric field of 0.3 kV/cm, which is roughly 60% or the RREA threshold and consistent with *Marshall and Stolzenburg* [2001]. Although the vertical length of the leader and ambient electric field strengths only depend on the total charge in the charge layers and the vertical separation between them, the threshold field strengths ($E_{max}$, $E_{th}$, and $E_{RREA}$) scale with density and pressure. To accurately scale the threshold field strengths from sea level to the upper altitude of the leader, we use the atmospheric density profile from the MSIS database (http://omniweb.gsfc.nasa.gov/vitmo/msis_vitmo.html); the scaling can be described by

$$E(h) = E(0)\frac{n(h)}{n(0)}, \qquad (9)$$

where $n(h)$ is the density at the upper altitude, $h$, of the leader channel.

**Table 2.** The Upper Altitude, $h$, the Vertical Length, $L$, the Ambient Electric Field Strength, $E_o$, the Ratio of the Ambient Electric Field to the RREA Threshold at the Upper Altitude, $E_o/E_{RREA}(h)$, and the Corresponding Potential Difference From the Center to the Tips of the Leader $\Delta U_{tip}$ (equation (7))[a]

|  | $h$ (km) | $L$ (km) | $E_o$ (kV/cm) | $E_o/E_{RREA}$ (h) | $\Delta U$ (MV) |
|---|---|---|---|---|---|
| *Rakov and Uman* [2003] | 11 | 6 | 0.2/0.4 | 0.26/0.53 | 120/240 |
| *Rakov and Uman* [2003] | 12 | 5 | 0.2/0.4 | 0.31/0.62 | 100/200 |
| *Marshall and Stolzenburg* [2001] | 9.9 | 3.4 | 0.38 | 0.42 | 65 |
| *Lu et al.* [2010] and *Shao et al.* [2010] | 14 | 5.5 | 0.3 | 0.64 | 165 |

[a]The values are derived from the references in the first column.

For each of these cases, we estimate the sum of the ambient electric field and the leader electric field and determine the following parameters: (1) the threshold field strengths, $E_{max} = 50$ kV/cm, $E_{th} = 32$ kV/cm, and $E_{RREA} = 2.86$ kV/cm, at sea level density and pressure, scaled to the upper altitude of the leader channel (equation (9)); (2) the distances from the tip of the leader channel to where the electric field drops below the threshold field strengths; here $l_{max}$ is the distance to $E_{max}$, $l_{th}$ to $E_{th}$, and $l_{RREA}$ to $E_{RREA}$; and (3) the potential differences that correspond to the regions between these values, where $\Delta U_o$ for $0 < l < l_{max}$, $\Delta U_1$ for $l_{max} < l < l_{th}$, and $\Delta U_2$ for $l_{th} < l < l_{RREA}$ (equation (7)).

From these values we determine (1) the potential difference in the AM region, $\Delta U_{AM} = \Delta U_1 + \Delta U_2$; (2) the total potential difference from the tip of the leader to the end of the AM region, $\Delta U_{tot} = \Delta U_o + \Delta U_{AM}$; and (3) the total number of avalanche lengths $N_\lambda$ in the region from $l_{th}$ to $l_{RREA}$ (equation (6)).

The results are shown in Figure 4. In Figure 4 (top), we present the potential differences in the AM region $\Delta U_{AM}$. In Figure 4 (middle), we present the ratio $\Delta U_{AM}/\Delta U_{tot}$, and in Figure 4 (bottom), we show the number of avalanche lengths in the AM region $N_\lambda$. These are plotted as a function of the potential difference from the tip of the leader to the end of the AM region, $\Delta U_{tot}$ (equation (6)).

Such potential calculations assume an isolated straight channel, but in natural lightning this may not be the case. Therefore, we also introduce the effect of horizontal development and branching of the leader channel in the negative charge region. The lightning leader is typically initiated close to the main negative charge region where the charge density is higher and therefore the electric field is stronger. While the negative end of the leader is directed away from the negative charge region, the positive end develops within the negative charge region. If the positive channel develops horizontally and branches within the negative charge region, the channel may be approximated as a perfectly (or partly) conducting surface and thus acts as a mirror channel to the negative end. Using the method of images, one can show that the resulting electric field can be calculated as if the channel is twice the actual vertical extension. The potential difference over the entire length of the leader then becomes correspondingly larger. This process was also introduced in *Mallios et al.* [2013]. To consider this potential difference to be doubled is an extreme assumption and can be considered to be the upper limit. The actual potential difference ahead of the negative end of the leader will be less than doubled as the branched channels at the positive end of the leader never form a perfectly conducting plane. The results in Figure 5 show the effect of taking into account horizontal development of the positive end of the leader.

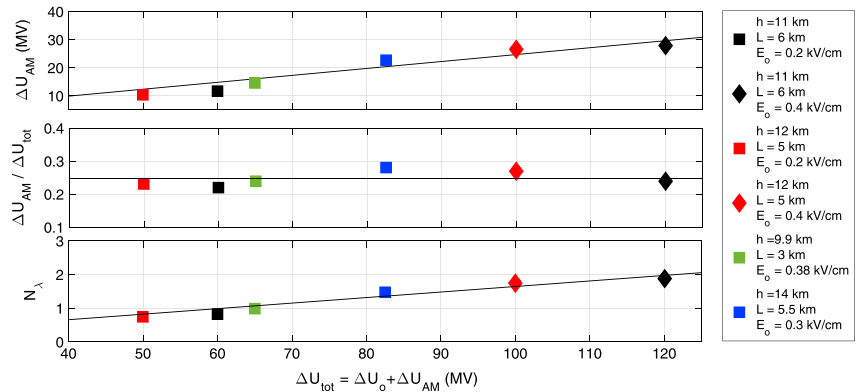

**Figure 4.** Results for each of the initial conditions given in Table 2. (top) The potential difference in the AM region $\Delta U_{AM}$. (middle) The ratio between the potential differences in the AM region and the sum of the potential difference in the high-field region and the AM region, $\Delta U_{AM}/\Delta U_{tot}$. (bottom) The number of avalanche lengths, $N_\lambda$, in the AM region. The results are plotted as a function of $\Delta U_{tot}$. The black lines in Figure 4 (top and bottom) correspond to the best fit linear function of the data points. The slope of the function in Figure 4 (top) coincides with the mean value of the data points in Figure 4 (middle), which is also plotted as a black line.

## 6. Discussion

### 6.1. The Relative Importance of the Ambient Electric Field

From balloon soundings one can see that the ambient electric field is less dependent on density and altitude but is related to the amount of charge in the charge regions [*Stolzenburg et al.*, 2007]. Thus, the effect of the ambient electric field becomes more important with increasing altitude and must be taken into account. To illustrate the significance of the ambient electric field, we estimate the ratio of the ambient potential difference in the AM region, $\Delta U_{amb}$, versus the total potential difference in the AM region, $\Delta U_{AM}$, that is, the second (right side) and first (left side) terms of equation (7), respectively. The results are plotted in Figure 6, for leader lengths of 3 km and 6 km and an ambient electric field strength of 0.4 kV/cm. This field strength constitutes between 30% and 99% of the RREA threshold (lower *x* axis) at assumed upper altitudes of the leader (upper *x* axis) from 7 km to 15 km. With these results, we find that for a given length, the relative importance of the ambient electric field only depends on the ratio $E_o/E_{RREA}(h)$. However, with increasing length of the leader, the ratio, $\Delta U_{amb}/\Delta U_{AM}$, decreases as the leader electric field becomes more important, which is expected. For ratios of $E_o/E_{RREA}(h)$ below 0.7, the relative importance of the ambient electric field is near constant. It is also interesting that at altitudes where the ambient electric field strength constitutes more than roughly 96% of the RREA threshold, the ambient electric field contributes more than the leader electric field to the potential difference in the AM region. Such fields can exist in local regions of a few hundred meters scale in the thundercloud, however, not over the full length of developed leader channels [*Stolzenburg et al.*, 2007]. It is clear that the role of the ambient electric field becomes increasingly important at higher altitudes.

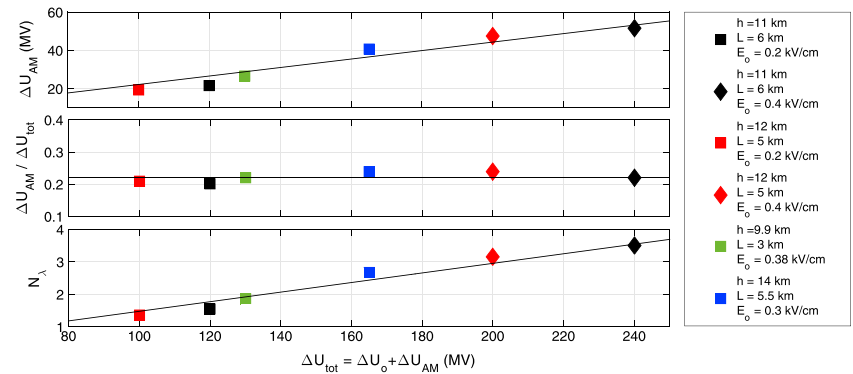

**Figure 5.** Results for each of the initial conditions given in Table 2 when taking into account the effect of horizontal development of the positive end of the leader. (top) The potential difference in the AM region $\Delta U_{AM}$. (middle) The ratio between the potential differences in the AM region and the sum of the potential difference in the high-field region and the AM region, $\Delta U_{AM}/\Delta U_{tot}$. (bottom) The number of avalanche lengths, $N_\lambda$, in the AM region. The results are plotted as a function of $\Delta U_{tot}$. The black lines correspond to the best fit linear functions in Figure 5 (top and bottom). The black lines in Figure 5 (top and bottom) correspond to the best fit linear function of the data points. The slope of the function in Figure 5 (top) coincides with the mean value of the data points in Figure 5 (middle), which is also plotted as a black line.

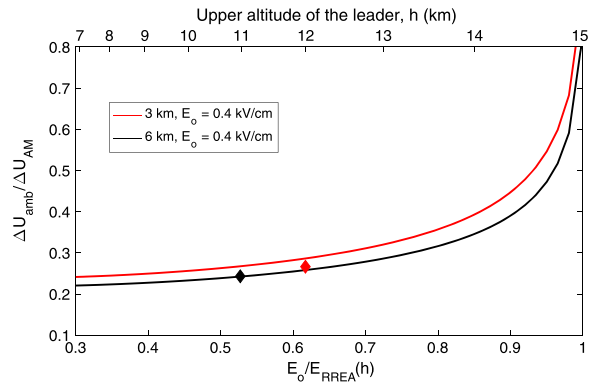

**Figure 6.** The ratio of $\Delta U_{amb}/\Delta U_{AM}$ as a function of $E_o/E_{RREA}(h)$, where $h$ is the upper altitude of the leader shown on the upper $x$ axis, for leader lengths of 3 km and 6 km in an ambient electric field of $E_o = 0.4$ kV/cm. The data points correspond to the first (black) and second (red) rows of Table 2, for an ambient electric field of 0.4 kV/cm. Note that we use the same symbol and colors as for the values plotted in Figure 4.

### 6.2. Potential Differences and Number of Avalanche Lengths in the AM Region

In Monte Carlo models, the threshold field strengths are given from the density used in the simulations. However, the estimated electric field only depends on the geometry of the leader channel and the assumed ambient electric field (equation (6)). This is critical when choosing how close to the leader tip the seed electrons can be initiated. Some models have made the strong assumption that the seed electrons can be initiated in the very high field region, denoted $\Delta U_o$ in Figure 1. At the distance of initiation, we have calculated the electric field to be roughly 8 times (at 30 cm) in *Xu et al.* [2012], 74 times (30 cm) in *Köhn and Ebert* [2015], and 60 times (15 cm) in *Celestin et al.* [2012], stronger than the conventional breakdown threshold. The potential difference in the AM region $\Delta U_{AM}$ was then 26.9 MV, 83.4 MV, and 145.0 MV, respectively. If we use the definition of the AM region as presented in this paper, and take into account scaling of the threshold field strengths, the corresponding potential differences become 17.7 MV, 26.6 MV, and 62.0 MV. *Köhn and Ebert* [2015] argue that high-energy electrons will be accelerated faster than the streamer zone can develop and that the effects of the streamer zone therefore can be neglected. This is only true if all the charges arrive at the tip of the channel instantaneously. To fully understand the maximum strength of the leader electric field and the initial position of the seed electrons, a better knowledge of the initial development of the streamer zone is necessary.

Figures 4 (top) and 5 (top) show that the potential difference in the AM region $\Delta U_{AM}$ ranges from ∼10 MV to 52 MV, depending on whether horizontal branching of the leader is taken into account. These results also show that the potential $\Delta U_{AM}$ has a linear dependence on the total potential $U_{tot}$ of the leader. In Figures 4 (middle) and 5 (middle), one can see that $\Delta U_{AM}$ is nearly constant at 24% and 22% of $\Delta U_{tot}$ in the two figures, respectively. While this ratio is expected to increase for very small potentials as the high-field region disappears, the nearly constant behavior covers the range of potentials roughly above 15 MV. This is in agreement with Figure 6, which shows the relative importance of the ambient field strength, for the given ratios of $E_o/E_{RREA}(h)$ (see Table 2). The potential difference $\Delta U_{AM}$ is directly related to the maximum energy that the accelerated electrons can achieve. Thus, a 52 MV potential difference can accelerate electrons to a maximum energy of 52 MeV, which can produce a much lower flux of photons that are equally energetic. However, due to the friction force experienced by the electrons, the energy of most of the electrons is expected to be less. This can be approximated by a fraction corresponding to the minimum friction force over the length of the AM region, that is $F_e = 2.0$ keV/cm at sea level [*Moss et al.*, 2006]. That gives a loss to friction of ∼14 MV over the AM region, which is $l_{max} − l_{RREA} \approx 270$ m at 11 km altitude. From these results, we can conclude that most of the electrons are expected to have energies below 38 MeV. Correspondingly, the photon energy distribution produced by these electrons will have energies weaker than that of the electrons. A rough estimate of the average energy gain can be derived from equation (2) [*Dwyer*, 2003; *Dwyer et al.*, 2012; *Skeltved et al.*, 2014]. An average energy of 7.3 MeV, or less, of avalanche length can then be expected. That corresponds to roughly 25 MeV. Thus, the candidates considered in this paper are of insufficient energy to explain the observed maximum energies of TGFs of at least 40 MeV [*Marisaldi et al.*, 2010]. Note that this result depends on the assumed initial position of the electrons (boundaries of the AM region).

To explain the photon intensity and energy distribution of TGFs, an electron energy distribution with an exponential cutoff at between 7.0 MeV and 7.3 MeV has been inferred [*Dwyer and Babich*, 2011; *Xu et al.*, 2012; *Skeltved et al.*, 2014]. This spectrum is typical for a fully developed RREA and is called the RREA spectrum. *Celestin et al.* [2015] used the same electric field constraints that have been implemented in this study but used sea level density and pressure. They showed that to reach a fully developed RREA spectrum, the potential difference from the center of the leader to its tip $\Delta U_{tip}$ must be approximately 300 MV. In *Dwyer and Babich* [2011], simulations in a homogeneous electric field showed that a steady state electron distribution can be obtained at less than 5 avalanche lengths. We have estimated the number of avalanche lengths $N_\lambda$ in the AM region (see Figures 4 (bottom) and 5 (bottom)). The values ranges from 0.7 to 3.5 depending on whether horizontal branching in the MN charge region is taken into account. The maximum potential difference between the lower and upper tips of the leaders considered in this study is 240 MV, which is less than that used by *Celestin et al.* [2015] and may be insufficient to produce a fully developed RREA spectrum. Thus, with our assumptions about $E_{max}$, $E_0$, length and upper altitude, and when horizontal branching of the leader is taken into account, only the maximum value of our results can be considered a candidate to obtain a fully developed RREA spectrum.

### 6.3. Can the Electric Field Ahead of Lightning Leaders Produce TGFs?

We have introduced new constraints and evaluated the leader electric field for assumptions derived from in situ measurements. From these results we will argue that if the energy loss due to friction is taken into account, these conditions are insufficient to explain a high flux of bremsstrahlung photons with energies of up to ~40 MeV, which is the observed maximum energy of TGFs [*Marisaldi et al.*, 2010]. By comparing our results to the results by *Celestin et al.* [2015], we find that only the maximum value of our results can be considered a candidate to obtain a fully developed RREA spectrum; this correspond to ~3.5 avalanche lengths in the AM region. To explain the maximum energies of TGFs, only leaders with potential differences between its tips of more than 500 MV can be considered. Note that this requires the assumption that horizontal branching of the leader channel in the MN charge region occurs. As far as we know, such large potential differences has not been observed.

Although we will argue that the maximum constraint on the electric field is realistic, Figures 4 (middle) and 5 (middle) show that with this constraint, only ~24 ± 3% (Figure 4) and 22 ± 2% (Figure 5) of the total potential difference from the tip of the leader to the end of the AM region $\Delta U_{tot}$ is within the AM region. Thus, 75%–80% is excluded. This may change if screening of the electric field, due to the effects of ionization during the leader step, is taken into account. If a larger fraction of $\Delta U_{tot}$ is within the AM region, the maximum energies of the electrons, and hence the bremsstrahlung photons, can become correspondingly larger.

Furthermore, most observed TGFs have energies below 40 MeV [*Smith et al.*, 2005; *Briggs et al.*, 2010; *Marisaldi et al.*, 2010] and some are within the energy range of the candidates considered in this paper. However, with our assumptions, these candidates cannot explain a fully developed RREA spectrum.

## 7. Conclusion

In this paper, we have modeled the electric field ahead of the leader tip and evaluated how it can accelerate and multiply high-energy electrons. We have used initial conditions that are consistent with electric field measurements of thunderclouds and estimates of the production altitudes of TGFs. Finally, we have argued that the potential difference ahead of a lightning leader can be increased by a factor of up to 2 due to horizontal development and branching of the positive end of the leader.

1. We have argued that the maximum electric field strength that can exist in the region ahead of the leader tip depends on how quickly the potential increases at that tip. As the timescale of the increase of the potential is not well understood, we set an upper limit to $E_{max} = 1.5 \cdot E_{th} \sim 50$ kV/cm. The lower limit is the electric field threshold required to sustain RREAs, $E_{RREA} = 2.86$ kV/cm, at sea level. Comparisons with existing models have shown that these constraints have important effects on the potential difference, and therefore the number of avalanche lengths, in the AM region.

2. By relating the boundary conditions of the AM region to the threshold field strengths, we have shown that scaling the maximum field and RREA threshold is important. It determines how close to the leader tip seed electrons can be initiated and the extent of the AM region. This scaling is important to correctly estimate the potential difference in the AM region.

3. We have shown that the relative importance of the ambient field to the leader field, on their respective contributions to the potential drop in the AM region ($\Delta U_{amb}/\Delta U_{AM}$), is nearly constant for ratios below ~0.7 but increases significantly at higher altitudes. For a ratio of more than 0.96, the relative importance of the ambient electric field is higher than that of the leader electric field. We also show that the ratio depends slightly on the length of the leader (given the same ambient electric field). Although the relative importance of the ambient electric field, compared to the leader electric field, increases significantly, thermal acceleration of low-energy electrons to become seed electrons is still necessary to explain the production of TGFs.

4. Given our assumptions of $E_{max}$ and $E_o$, we find that the maximum potential difference in the AM region is roughly 52 MV corresponding to a maximum number of avalanche lengths of roughly $N_\lambda = 3.5$. This is obtained for a potential difference between the leader and the ambient potential at the leader tip ~240 MV (480 MV over the entire vertical length of the leader) and assumes horizontal branching of the channel in the MN charge region. These values are insufficient to explain the maximum photon energy associated with observed TGFs and close to the minimum required to produce a fully developed RREA spectrum.

## Appendix A: Method of Moments for an Electrostatic Problem

### A1. Basic Terms

The Poisson equation $\nabla^2\phi = -\rho/\epsilon_0$ may be represented as

$$\int G(\mathbf{r}, \mathbf{r}')\rho(\mathbf{r}')\,d^3\mathbf{r}' = \phi(\mathbf{r}), \qquad G(\mathbf{r}, \mathbf{r}') = \frac{1}{4\pi\epsilon_0|\mathbf{r} - \mathbf{r}'|} \tag{A1}$$

In some problems, such as a conductor placed in an external field, this equation has to be solved for $\rho$ with given $\phi$ and thus is a Fredholm integral equation of the first kind. It may be written symbolically as

$$\hat{G}\rho = \phi \tag{A2}$$

where $\hat{G}$ is a linear operator and $\rho$ and $\phi$ are functions.

The methods of moments is a method of solving (A1) or (A2) by discretization; i.e., $\hat{G}$ is represented as a matrix while $\rho$ and $\phi$ as vectors. We approximate the unknown function as a linear combination of *basis (expansion) functions* $u_i$:

$$\rho(\mathbf{r}) \approx \sum_i \rho_i u_i(\mathbf{r}) \tag{A3}$$

and try to satisfy (A2) with a discrete number of conditions

$$\phi_j = (w_j, \phi)$$

where $w_j$ are the *testing (weighting) functions* and the scalar product is defined as

$$(f, g) \equiv \int f(\mathbf{r})g(\mathbf{r})\,d\mathbf{r}$$

Substituting the discretized $\rho$, we get a system of linear equations

$$\sum_i G_{ji}\rho_i = \phi_j \tag{A4}$$

where $G_{ji} \equiv (w_j, \hat{G}u_i)$, which may be solved if matrix $G_{ji}$ is invertible.

### A2. Surface and "Axis" Discretization Algorithms

Let us consider a cylindrically symmetric conducting object in the cylindrical system of coordinates $\mathbf{r} = \{r_\perp, \theta, z\}$. The object surface is described parametrically

$$z = Z(\tau), \qquad r_\perp = R(\tau), \qquad \tau_{min} < \tau < \tau_{max}$$

In the problem solved in this paper, the external field is uniform and axial, $\phi = -E_0 z$, and therefore, due to symmetry, neither $\rho$ nor $\phi$ depend on the azimuthal angle $\theta$. Moreover, the charge is concentrated on the surface:

$$\rho(\mathbf{r}) = \int_{\tau_{min}}^{\tau_{max}} \lambda(\tau) S(\mathbf{r}, \tau) \, d\tau, \qquad S(\mathbf{r}, \tau) = \frac{\delta \left[ r_\perp - R(\tau) \right]}{2\pi R(\tau)} \delta[z - Z(\tau)]$$

where $\lambda$ is the linear charge density (in $\tau$) and $S$ describes the surface shape in terms of Dirac delta functions.

The basis (expansion) functions $u_i(\mathbf{r})$, $i = 1 \ldots N$ are represented in terms of basis functions $u_i^\tau(\tau)$ in $\tau$:

$$u_i(\mathbf{r}) = \int_{\tau_{min}}^{\tau_{max}} u_i^\tau(\tau) S(\mathbf{r}, \tau) \, d\tau$$

We take $u_i^\tau(\tau) = 1/\Delta \tau_i$ when $\tau \in (\tau_i, \tau_{i+1})$ and zero otherwise ("pulse" functions), with $\Delta \tau_i = \tau_{i+1} - \tau_i$, $\tau_1 = \tau_{min}$, $\tau_{N+1} = \tau_{max}$.

Although $u_j(\mathbf{r})$ are the same in the surface and axis methods, the choice of the testing (weighting) functions $w_j$ is different. In the surface method we take $w_j \equiv u_j$. Thus, we evaluate the potential at the surface:

$$(u_j, \phi) = \frac{1}{\Delta \tau_i} \int_{\tau_i}^{\tau_{i+1}} \phi \left[ R(\tau), Z(\tau) \right] \, d\tau$$

In the axis method of *Balanis* [2012], however, $w_j \neq u_j$ as the potential is evaluated on the axis in the center of the $j$th surface element:

$$(w_j, \phi) = \phi \left( 0, Z_j \right)$$

where $Z_j$ is an average value of $Z(\tau)$ in $\tau \in (\tau_j, \tau_{j+1})$, i.e., $w_j(\mathbf{r}) = \delta(\mathbf{r}_\perp) \delta(z - Z_j)$.

The discretized operator $\hat{G}$ in the surface method is

$$G_{ji} = (u_j, \hat{G} u_i) = \int \frac{u_j(\mathbf{r}) u_i(\mathbf{r}')}{4\pi \epsilon_0 |\mathbf{r} - \mathbf{r}'|} \, d\mathbf{r} \, d\mathbf{r}' = \int d\tau_1 \int d\tau_2 \, u_j^\tau(\tau_1) u_i^\tau(\tau_2) G(\tau_1, \tau_2)$$

where

$$G(\tau_1, \tau_2) = \int \frac{S(\mathbf{r}, \tau_1) S(\mathbf{r}', \tau_2)}{4\pi \epsilon_0 |\mathbf{r} - \mathbf{r}'|} \, d\mathbf{r} \, d\mathbf{r}' =$$

$$\int_0^{2\pi} \frac{1}{4\pi \epsilon_0 \sqrt{\left[ Z(\tau_1) - Z(\tau_2) \right]^2 + \left[ R(\tau_1) - R(\tau_2) \cos \chi \right]^2 + \left[ R(\tau_2) \sin \chi \right]^2}} \frac{d\chi}{2\pi}$$

which evaluates to

$$G(\tau_1, \tau_2) = \frac{K(m)}{2\pi^2 \epsilon_0 D_+}, \quad m = \frac{4R(\tau_1)R(\tau_2)}{D_+^2} = 1 - \frac{D_-^2}{D_+^2}, \quad D_\pm^2 = \left[ Z(\tau_1) - Z(\tau_2) \right]^2 + \left[ R(\tau_1) \pm R(\tau_2) \right]^2$$

where $K(m)$ is the complete elliptic integral of the first kind. The second version of expression for $m$ may be used when elements are close to each other, i.e., $D_- \rightarrow 0$ so that $m \approx 1$. We substitute it into the above expression for $G_{ji}$ which is evaluated numerically. We note that $G(\tau_1, \tau_2)$ is infinite at $\tau_1 = \tau_2$ (because then $m = 1$), which occurs in calculation of diagonal elements $G_{ji}$, but the singularity can be integrated through. We may comment that avoiding singularity was the motivation of *Balanis* [2012] to use the axis testing (weighting) functions. However, this leads to a much more serious problem of numerical instability, which we will shortly discuss.

### A3. Discussion of Stability

Operator represented by Green's function $G$ in (A1) is symmetric and positive definite, which is demonstrated by evaluating it in wave vector $\mathbf{k}$ domain, where it is diagonal with values $\tilde{G}(\mathbf{k}) = \frac{1}{\epsilon_0 |\mathbf{k}|^2} > 0$ for all $\mathbf{k} \neq 0$. These properties mean that solving (A2) is equivalent to minimizing the functional of $\rho$

$$U[\rho] = \frac{1}{2}(\rho, \hat{G}\rho) - (\rho, \phi) \rightarrow \min \text{ over } \rho(\mathbf{r}) \tag{A5}$$

and the minimum exists.

After a discretization which is symmetric (i.e., the basis and testing functions are the same, $u_i \equiv w_i$), which is the case in the surface method, the discretized matrix $G_{ij}$ remains a positive-definite symmetric matrix and solving (A4) is still equivalent to a minimization problem (A5) but on a limited set of functions $\rho$ given by (A3). As this limiting becomes less and less restrictive when we decrease the element size, the minimum should get closer and closer to the unconstrained solution. The actual speed of convergence probably depends on how we choose our basis (expansion) functions $u_i$. A robust analysis of the convergence speed was beyond the efforts we were willing to spend on this paper.

In the axis method, however, the discretized operator $G_{ji}^{\text{axis}} = (w_j, \hat{G}u_i)$ is neither symmetric nor positive definite, and there is no analogous minimization problem. The method becomes unstable when element sizes are smaller than the radius of the conductor. This may be understood by the following reasoning. The oscillations in $\rho$ with wavelength $l \ll R$ (which can be as small as the element size, $l_{\min} = 2\Delta Z$) on the surface create only very small oscillations in $\phi$ on the axis, with suppression factor $\sim e^{-2\pi R/l}$. This means that when we solve for $\rho$, very small fluctuations of $\phi$ on the axis are amplified by a large factor $e^{2\pi R/l} \sim e^{\pi R/\Delta Z}$. The inability to use small elements leads, among other things, to inability to accurately calculate fields close to the conductor.

**Acknowledgments**

This study was supported by the European Research Council under the European Union's Seventh Framework Programme (FP7/2007-2013)/ERC grant agreement 320839 and the Research Council of Norway under contracts 208028/F50, 216872/F50, and 223252/F50 (CoE). The data used in this paper are generated by the use of the surface method of moments as described in the appendix, with the specified conditions from the main body of the paper.